# Evaluating Massive MIMO Precoding based on 3D-Channel Measurements with a Spider Antenna

M. Arnold, M. Gauger, S. ten Brink
Institute of Telecommunications, Pfaffenwaldring 47, University of Stuttgart, 70569 Stuttgart, Germany
Email: {arnold,gauger,tenbrink}@inue.uni-stuttgart.de

***Abstract*—Massive Multiple-Input Multiple-Output (MIMO) communications uses a large number of antennas at the base station to increase the data rate and user density in future wireless systems. For simulation, it has become common practice to use i.i.d. complex Gaussian matrix entries to obtain an average MIMO channel behavior. More refined models have been devised and proposed to standardization bodies; yet, channel modeling remains an active area of research, as current models tend to be, still, quite limited, e.g., when it comes to evaluating clustering algorithms, with regions of spatial orthogonality for concurrent scheduling of users, which is an essential concept in massive MIMO precoding. For this, spatial correlations need to be included. To further refine channel modeling, we have built a "spider antenna" prototype that allows spatially continuous measurements in three dimensions, enabling a high-resolution channel sampling over, initially, a volume of 2m x 2m x 2m for indoor measurements. Several experiments have been conducted to illustrate the new insights to be gained when studying user orthogonality, clustering and precoding in a massive MIMO context. Furthermore, the influence of antenna array geometry and user spacing on the achievable rate over actually measured channels is studied.**

## I. Introduction

With decreasing hardware cost and increasing demand for high data rates, massive MIMO has become a promising candidate for future wireless communications [1], [2], [3]. An anticipated over-provisioning of antennas can be used for spatially orthogonalizing users in space. With "massive" MIMO, base stations employ a large number of antennas (e.g., 64 and more) while mobile terminals typically are assumed to have only one, or a few antenna(s). This is expected to lead to an increased sum rate of the cellular system [4]. For downlink precoding, channel estimates can be obtained at the base station based on up-link training while exploiting channel reciprocity in time division duplex (TDD) mode.

One particularly interesting aspect of massive MIMO research considers channel modeling. Currently, the most widely used model in simulation of cellular systems assumes a channel matrix with independent and identically distributed (i.i.d.) complex Gaussian entries [5]. This model does not contain any specific spatial modeling or correlation. Therefore, an interesting research topic is to find parametrized models that better represent the detailed spatial correlations in a massive MIMO context.

The influence of precoding in a spatially fixed environment was investigated in [6], [7]. To define a parametrized true 3D-channel model which includes a spatial component, the first step is to create clusters from the measured area [8], [9], [10]. Therefore, an area measurement can visualize the energy distribution over the area (or volume) allowing to better study spatial correlations. Moreover, it is possible to observe wavefronts of single antenna links to verify the time stability of the measurement. By interpreting the area measurements as an indoor access point ("basestation"), different antenna geometries can be used to study user correlation and to evaluate achievable rates for different precoding schemes.

The paper is organized as follows: Section II describes the system model and outlines the experimental set-up. Section III explains the calibration procedures and sanity checks, to justify the insights gained. In Section IV, the influence of antenna geometry and precoding method on the achievable channel rates are investigated. Finally, Section V renders some conclusions.

## II. Experimental Set-up

### *A. Area Measurements using a Spider Antenna*

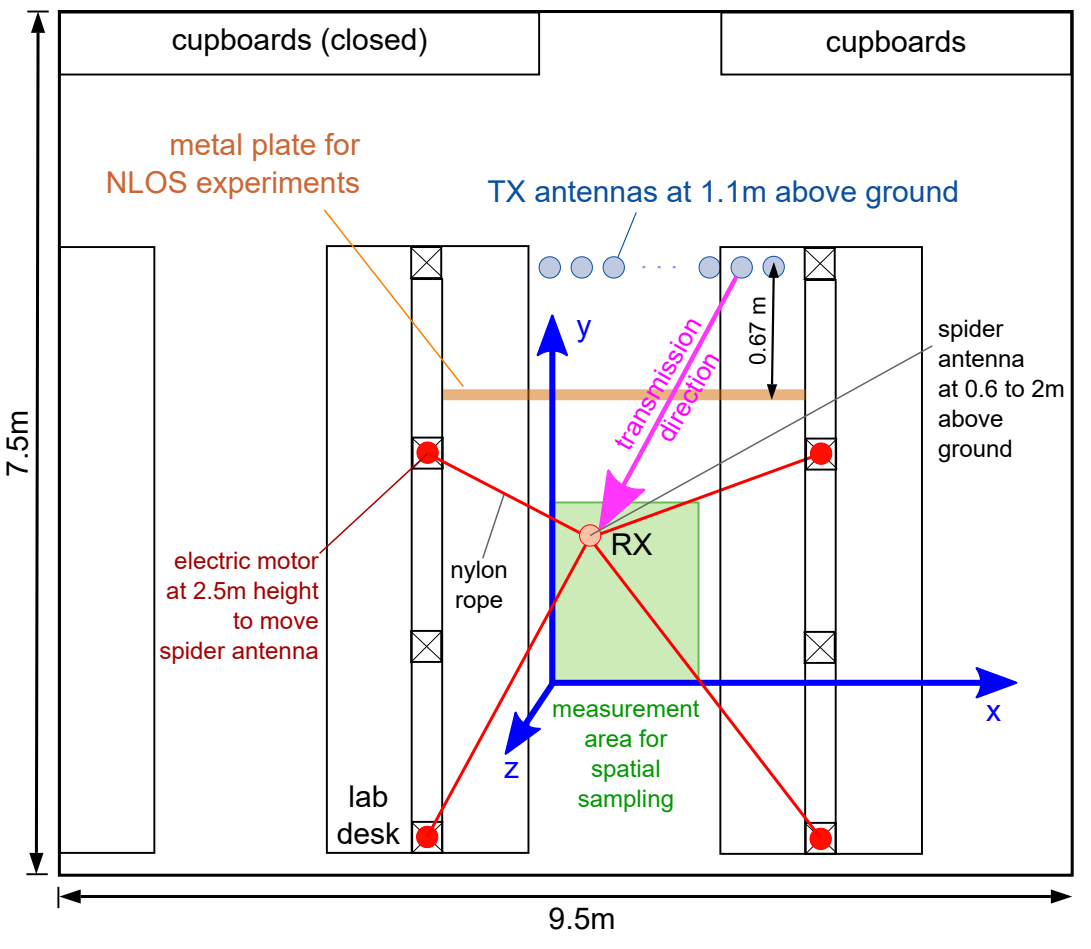

Fig. 1. Area measurement set-up with a spider antenna.

The spider antenna is based on four electric motors, and two software defined radios (SDR): The receiving SDR (USRP B200 from "Ettus Research") is connected to the four motors by nylon ropes, as illustrated in Fig. 1; by this, an arbitrary path in three dimensions can be traced. The transmitting SDR is a USRP B210, which is placed along a line parallel to the measurement area. In this configuration the setup can measure within a volume of up to $2\,\mathrm{m} \times 2\,\mathrm{m} \times 1.8\,\mathrm{m}$. An area is spatially scanned by tracing a meander-like path with a resolution of $1\text{mm}$ in $x$, and $4\text{cm}$ in $y$ and $z$-direction, respectively. This kind of spatial sampling was done so that adjacent measurement points are closer than $\lambda/2 \approx 6\text{cm}$. To measure a complete area (e.g., the $x$, $y$-plane) takes about 10 minutes. As the movement is rather slow, with a constant speed at $0.178\text{m/s}$, Doppler effects can be neglected.

### B. Channel Sounding

The channel measurement was carried out in the ham radio frequency band at $2.35\text{GHz}$, with a maximally allowed transmit power of $13.15\text{dBm}$ ERP. Therefore it is possible to measure the channel at a bandwidth of $40\text{MHz}$ without interference caused by WiFi transmissions. The antennas on both sides (TX/RX) are dipole antennas (vertical polarization). For channel estimation, an OFDM signal with 1024 subcarriers and BPSK modulation per subcarrier was used. A cyclic prefix taking $25\%$ of the symbol duration is inserted for multipath mitigation; $90\%$ of the subcarriers were used for channel sounding, with a subcarrier spacing of 39kHz. Packets of 20 OFDM symbols each are transmitted every $5\text{ms}$, providing a gain of $10\log_{10} 20 \approx 13\text{dB}$ for channel estimation by averaging. The noise was estimated by observing the unused subcarriers. For each measurement it was verified that the SNR stayed above $30\,\text{dB}$. The USRPs were synchronized with a GPSDO (GPS disciplined oscillator) signal. Moreover, both, RX and TX are driven by the same PPS (pulse per second) and $10\text{MHz}$ reference clock, to ensure that there is no phase or amplitude drift. Note that this specific setup measures SISO channels only; however, for our static environment (indoor, lab conditions) several measurement results can be combined (“overlaid”) to obtain the corresponding MIMO channel scenario, as is further verified in Section III. This technique was also successfully applied in, e.g., [11].

## III. Measurement Verifications

### A. TX and RX Phase Stability

Several long time measurements in fixed Line-of-Sight (LoS) and Non-Line-of-Sight (NLoS) scenarios were carried out to verify that the channel sounding equipment does not influence the channel, and that the measurement results are stable and reproducible. For the channel to be stable there must not be any phase or amplitude drift over time. A phase shift $\Delta\phi$ over time would show up as a shift in position by $\Delta d_{\text{off}} = \lambda \cdot \Delta\phi / 2\pi$.

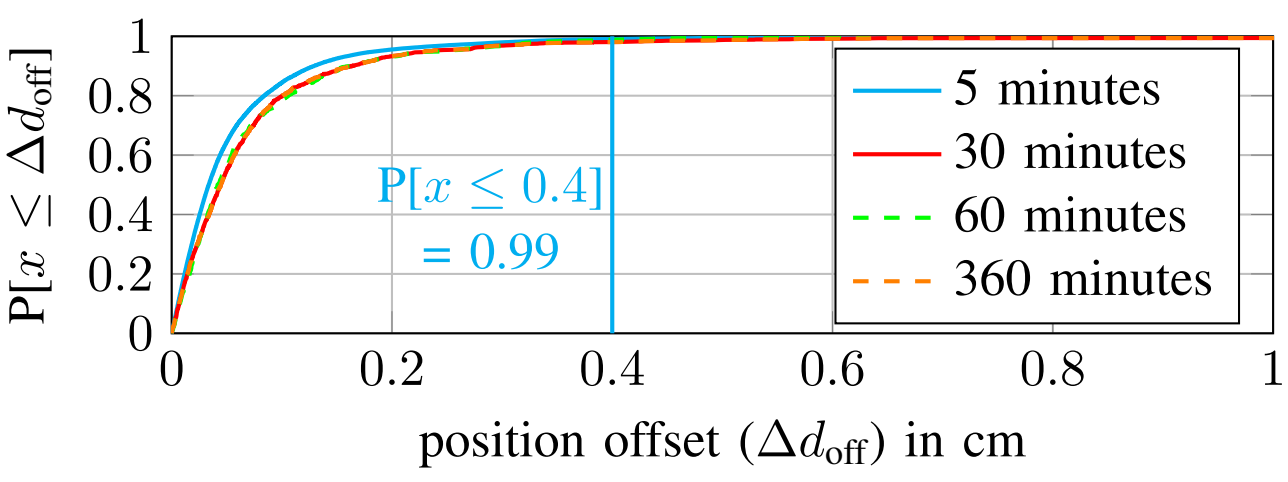

Fig. 2. Cumulative distribution function (CDF) of position offsets.

Fig. 2 shows that $99\%$ of the position offsets of successive measurements stay below $4\text{mm}$. As can be seen, the channel hardly changes within the first 30 minutes, and thus, stays virtually constant over 6 hours under lab conditions, allowing to perform several area measurements with different TX antenna positions consecutive in time. Note that, for this measurement, only the result for the first subcarrier of the OFDM signal was evaluated and plotted; yet, it was verified that the other subcarriers behave accordingly.

### B. Ensuring Reproducibility of Measurements

To validate the experimental setup it is, again, important to show that the measurement is reproducible. For this, two area measurements with the exact same scenario, and one area measurement where the transmitter was moved to a different position, were compared against each other.

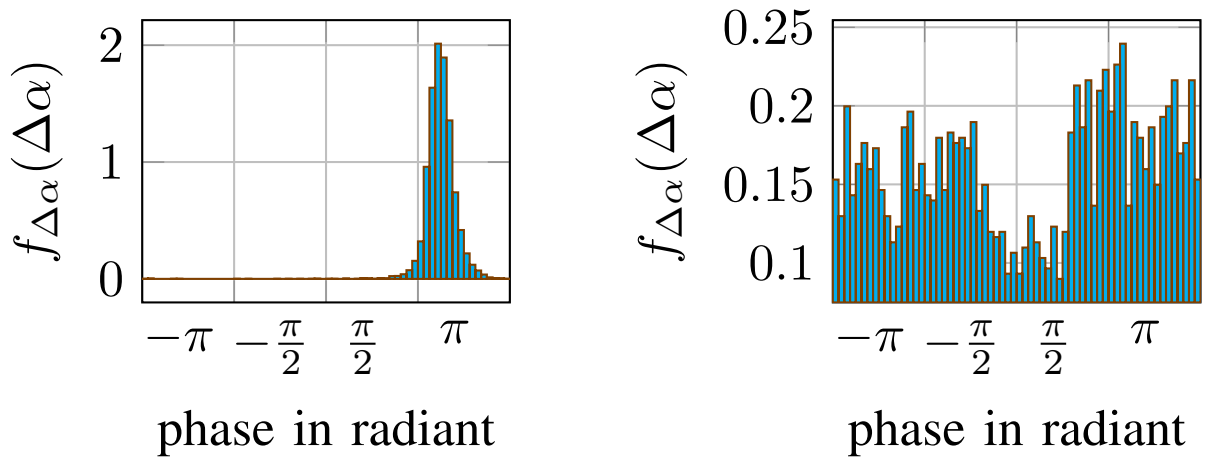

(a) TX antenna position unchanged. (b) Different TX antenna position.

Fig. 3. Probability density function (PDF) of the phase difference $\Delta\alpha$ between two consecutive area (i.e., plane) measurements.

Two area planes were measured consecutively, and the probability density function (PDF) of the phase difference $\Delta\alpha$ per spatial coordinate is shown in Fig. 3. As can be seen, if the the TX antenna position is not changed, the phase difference is concentrated around a mean value, exhibiting a Gaussian-like distribution, indicating a rather small variation between the two measurements. After shifting the TX antenna position, however, by a distance of $\lambda$ in $x$-direction, the phase difference becomes quite arbitrary, and more close to a uniform distribution. From this we can, as before, infer that the area plane measurements are reproducible, and thus, under lab conditions, can be carried out consecutively.

### C. Verification of Line-of-Sight (LoS) Measurements

To verify the physical plausibility, we first resort to LoS measurements as they can be handled mathematically quite conveniently. The (baseband) phase and amplitude of an LoS-channel can be computed using

$$h_{\text{sim}} = \frac{\lambda}{4\pi r} e^{j2\pi \frac{r}{\lambda_{\text{sub}}}} \tag{1}$$

where $r$ is the distance between TX and RX, and $\lambda_{\text{sub}}$ is the wavelength of the respective subcarrier considered.

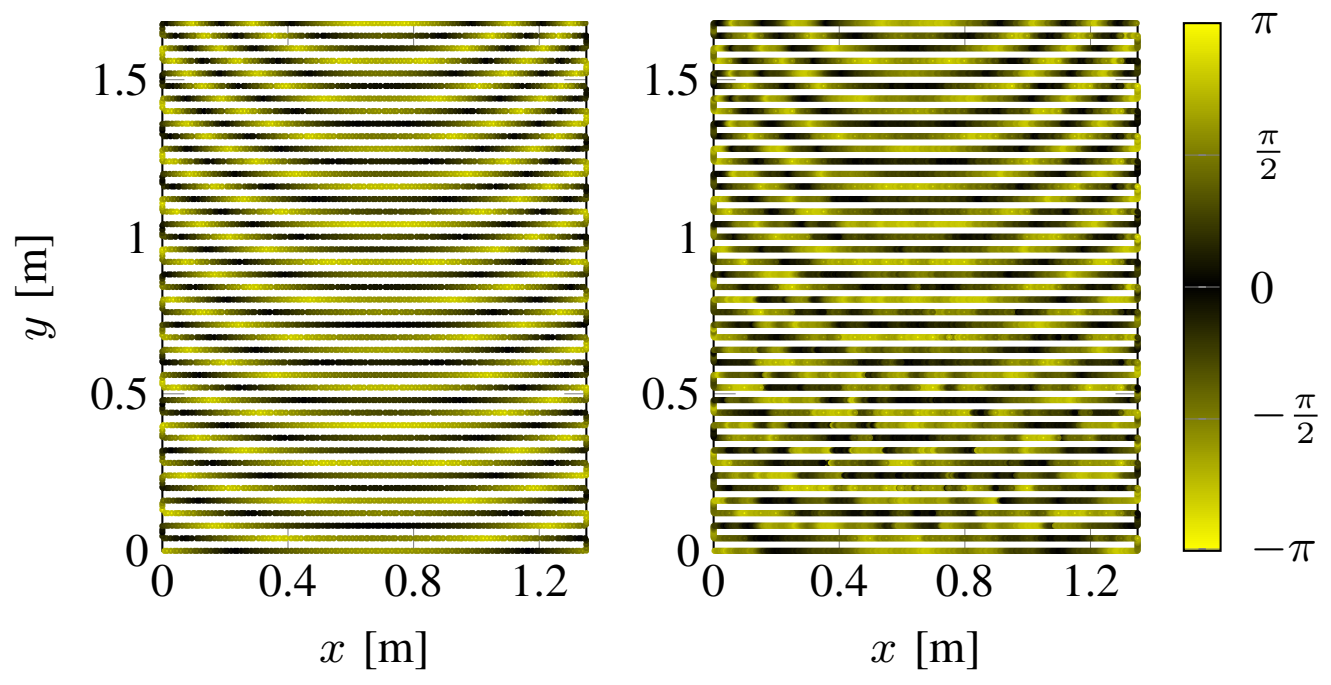

(a) Simulated (ideal LoS). (b) Actually measured.

Fig. 4. "Wavefront images", simulated vs actually measured baseband phase in the $x$, $y$-plane; TX and RX are on same height at 1.1m above ground.

Fig. 4 compares the baseband phase of the measured LoS scenario (right) with the simulation model according to (1) in the $x$, $y$-plane, using the set-up of Fig. 1. Note that the resulting "wavefront images" match remarkably well. Moreover, the meander structure of the right figure illustrates the actual path that the spider antenna was tracing in $x$, $y$-space. The radial distance between any two rings corresponds to the wavelength $\lambda \approx 12.7\text{cm}$, as TX and RX are located on the same height. To further show the potential of the spider antenna for 3D channel studies, an $x$, $z$-plane measurement was also performed, as depicted underneath.

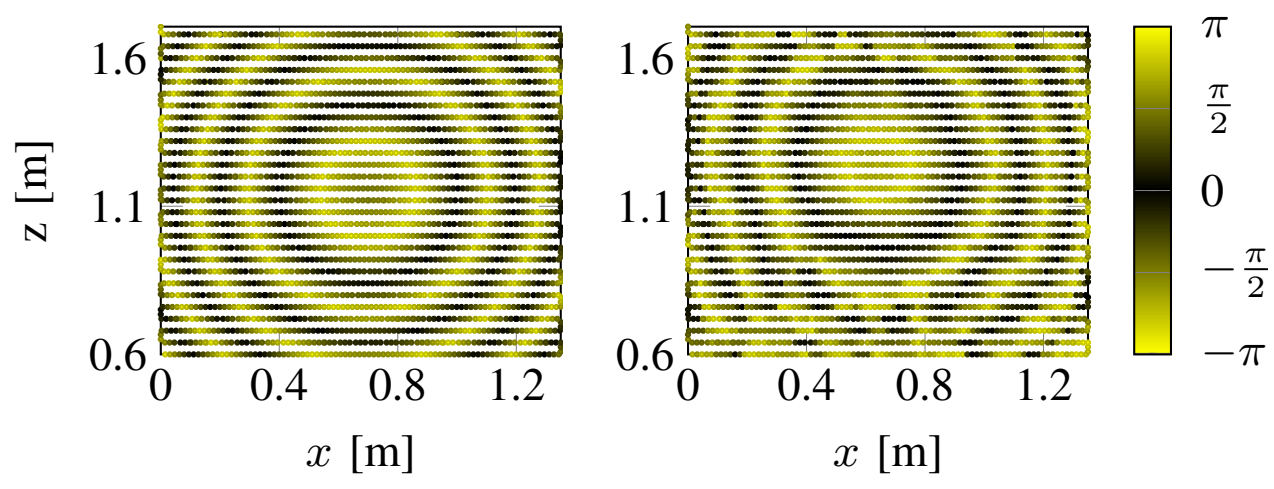

(a) Simulated. (b) Measured.

Fig. 5. Wavefront images, simulated vs actually measured baseband phase in the $x$, $z$-plane; TX and RX are 50cm apart in $y$-direction.

Again, Fig. 5 shows a remarkably good agreement of the wavefront measurements with the anticipated, theoretical LoS behavior, illustrating how the spider antenna is capable of scanning arbitrary 3D-structures in space for more comprehensive characterization of MIMO channels.

Recall that, for NLoS channels, no simple simulation model is available, unless sophisticated ray-tracing methods are employed that also take into account reflection properties of surfaces, and several other wave propagation effects. In such cases, the spider antenna is a convenient tool to capture spatially contiguous snapshots (area, or volume images of channel coefficients) of the actual channel behavior.

*From single antenna to a multi-antenna linear array*

For channel characterization, a 16-antenna linear array is arranged on a line parallel to the measurement path. Both, LoS as well as NLoS scenarios were measured (compare to Fig. 1). Different NLoS scenarios can be created by gradually inserting a tall metal plate in front of the linear array, as described in more detail later in Section IV-B. The individual array element positions are spaced $\lambda$ apart from each other, on a line at $y = 3.25\text{m}$. The LoS scenario can be compared to the theoretical model of (1). The received signal at each coordinate in the measurement area can be written as

$$\mathbf{y} = \mathbf{HPs} \tag{2}$$

where $\mathbf{H}$ is an $M \times N$ matrix (corresponding to $M$ receivers, i.e., coordinate points in area, and $N$ transmitters), $\mathbf{P}$ the precoding matrix, and $\mathbf{s}$ the transmit symbol vector. The Maximum-Ratio (MR) precoding matrix is given as $\mathbf{P}_{\text{MRC}} = \frac{1}{\sqrt{M}}\mathbf{H}^H$ (e.g., see [12]).

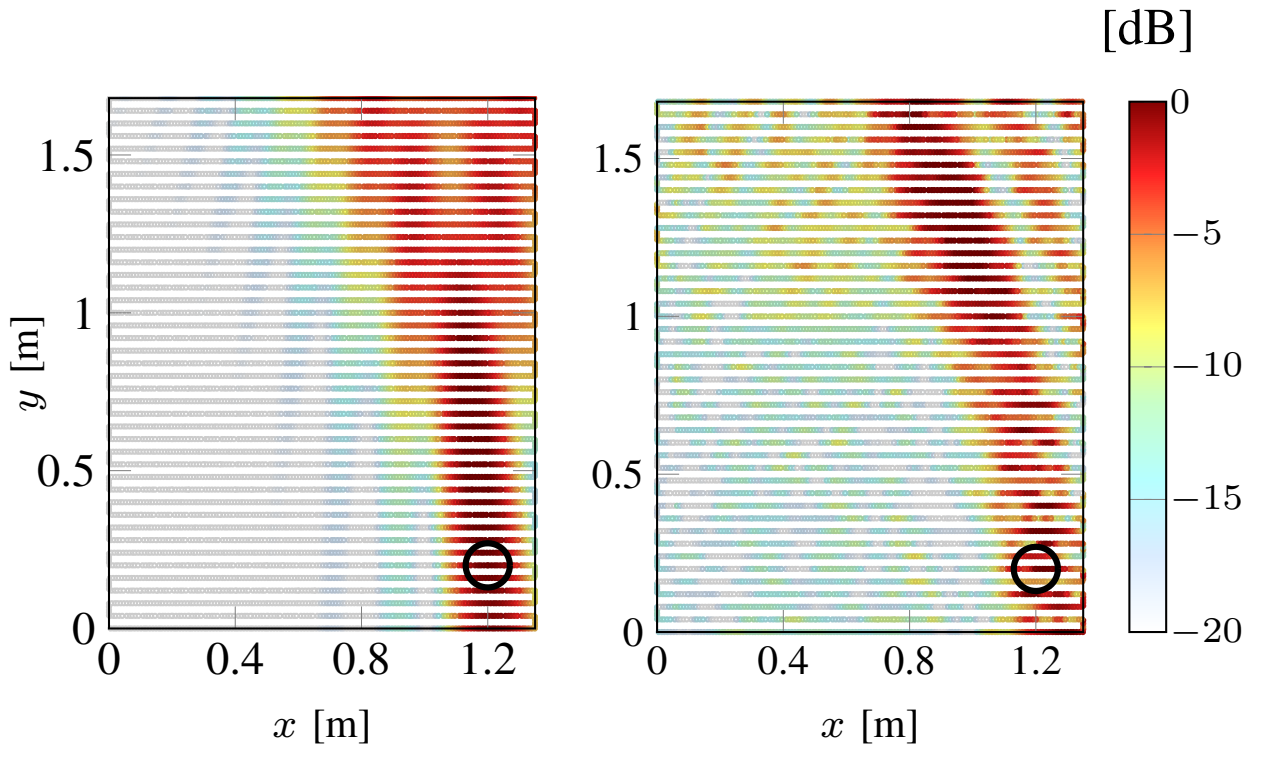

(a) Simulated (ideal LoS). (b) Measured.

Fig. 6. Spatial energy map, simulation vs spatial area measurement for MR precoding with respect to a target user at $x = 1.2\text{m}$, $y = 0.2\text{m}$, 16-antenna linear array, LoS scenario.

Fig. 6 shows, for each coordinate in the $x$, $y$-plane, the received energy when using MR precoding, with a target (user) position at $x = 1.2\text{m}$, $y = 0.2\text{m}$. The MR precoding results in a beam, which can be steered by the position to precode on. With increasing number of antennas the sidelobes vanish (for an illustration of this effect, see, e.g., webdemo [13]).Again, simulation and measurement of the spatial energy map are in remarkably good agreement, verifying the plausibility of the spider antenna measurements. We are now ready to move on to possible applications of the set-up, including some quantitative performance evaluation of different antenna geometries and scenarios.

## IV. Applications of Area Measurements

### *A. Studying clusters ("outside in"-scenario)*

Finding spatially contiguous "patches" of area (or volume) with similar or same propagation conditions allows to partition users into groups of quasi-orthogonal clusters within a coverage area; this knowledge can then be exploited for scheduling to increase the overall rate of the system. For such cluster modeling one typically has to rely on ray-tracing (e.g [8]). Area measurements enable testing of clustering algorithms against the actual channel behavior. As a first simple idea of quantifying the opportunity for building clusters, we define a spatial SIR measure and apply MR precoding with respect to a specific target position $(x_0, y_0)$: The desired signal is considered to be at $(x_0, y_0)$ while the interference is taken as the average power over *all other* positions $(x, y)$ in the measurement area. Using (2), this SIR-measure is calculated as

$$\overline{\text{SIR}}_{x_0,y_0} = \frac{\left|\mathbf{h}_{x_0 y_0}\mathbf{h}^{\text{H}}_{x_0 y_0}\right|^2}{\frac{1}{M-K}\sum\limits_{x,y}^{M-K}\left|\mathbf{h}_{xy}\mathbf{h}^{\text{H}}_{x_0 y_0}\right|^2} \quad (3)$$

where $\mathbf{h}_{xy}$ is the $1 \times N$-channel vector from the $N$ RX antennas to a position $(x,\, y)$ outside a circle with radius $\lambda/2$ around the target position $(x_0,\, y_0)$; $M$ is the total number of measured area points, and $K$ is the number of area points inside the circle. Studying the effect of MR precoding in the transition from LoS to NLoS scenarios appears to be particularly interesting; for this, a tall metal plate was inserted in between the TX and RX antennas (see Fig. 1) in four different positions: 1) no plate, 2) metal plate covering 1/3 of the RX antennas, referred to as "1/3 NLoS", 3) correspondingly, "2/3 NLoS", and a full NLoS scenario, where the plate covered all RX antennas, respectively.

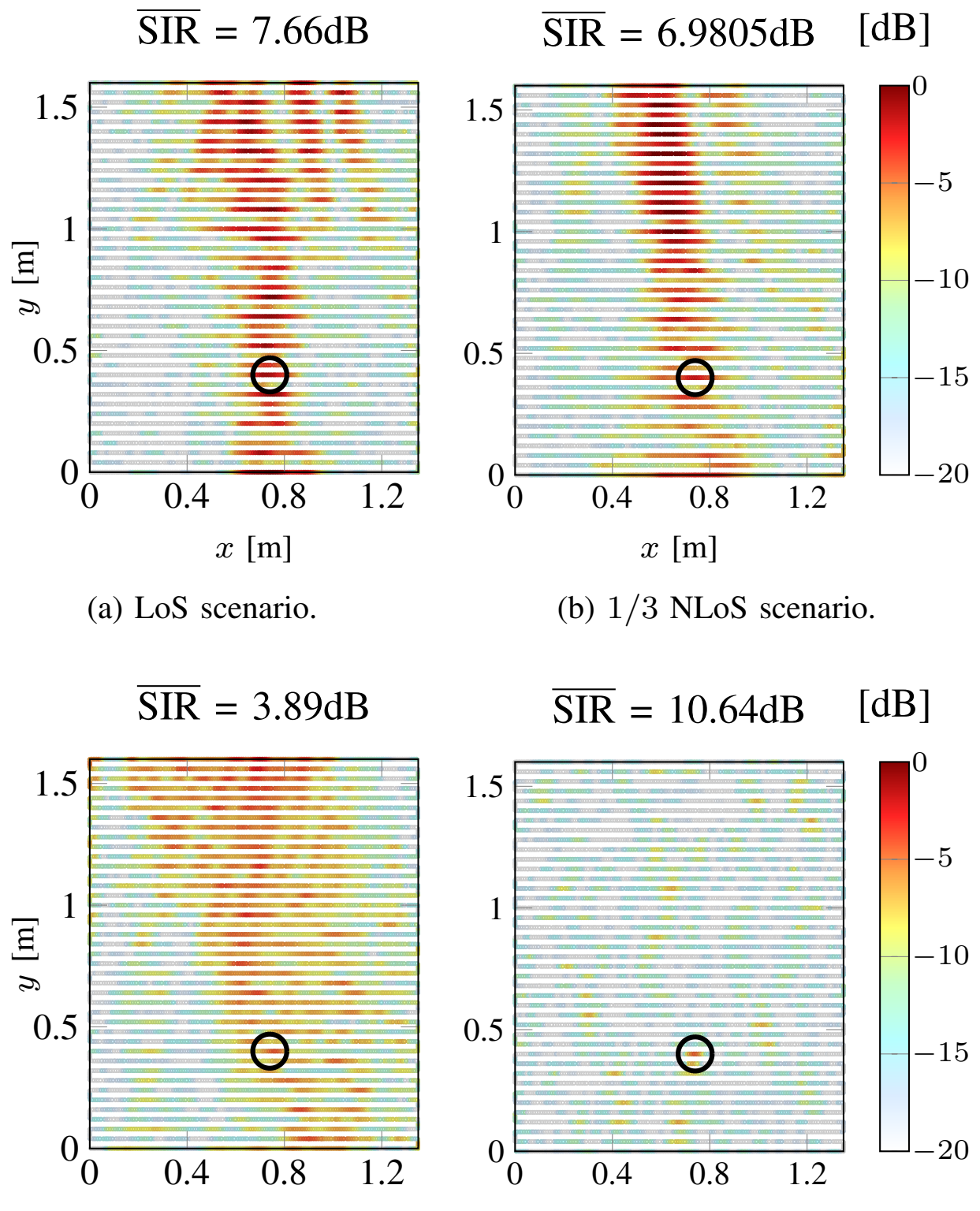

(a) LoS scenario. (b) 1/3 NLoS scenario.

(c) 2/3 NLoS scenario. (d) NLoS scenario.

Fig. 7. Spatial energy maps, transition from LoS to NLoS, MR precoding for target position at $x = 0.74$m, $y = 0.4$m, 16-antenna linear array.

Fig. 7 shows the energy distribution over the measurement area when MR precoding is applied to, e.g., the position at $x = 0.74$m, $y = 0.4$m. By comparing the top left with the bottom right plot, it can be seen that MR precoding in NLoS gains 3dB over the LoS scenario in terms of the $\overline{\text{SIR}}_{x_0,y_0}$-measure defined in (3), owing to the increased diversity. For a fair visual comparison within the same color bar scaling, the plots were normalized to the maximum received energy of the LoS scenario, corresponding to 0dB, and the TX power was kept constant for all four experiments. Interestingly, in the transition from LoS to NLoS, the beamwidth first increases in the $1/3$ and $2/3$ NLoS case, leading to worse $\overline{\text{SIR}}_{x_0,y_0}$; the concept of "beamforming" completely dissolves in the full NLoS case, yielding, finally, an improved $\overline{\text{SIR}}_{x_0,y_0}$, even higher than that for the LoS case.

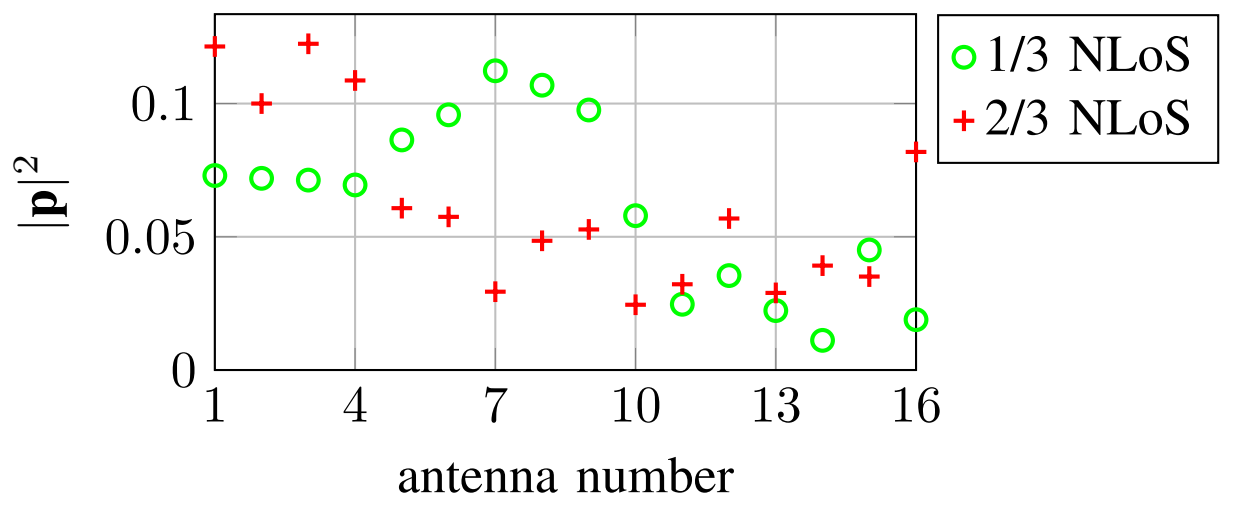

Fig. 8. Precoding energy at the different antenna positions within linear array.

Fig. 8 shows how MR precoding increases the transmit power $|\mathbf{p}|^2$ at those antennas not yet covered by the metal plate, effectively reducing the array size, so that, e.g., in the 2/3 NLoS scenario, only 4 antennas are effectively used. In the LoS as well as the full NLoS scenario, all antennas are used with equal weight, helping to lift the NLoS SIR-measure above that of the 2/3 NLoS scenario.

Obviously, with this kind of area measurement data, more sophisticated clustering measures as well as approaches, like the $k$-means algorithm, can be studied (e.g. [14]).

### *B. Channel rates and orthogonality ("inside out"-scenario)*

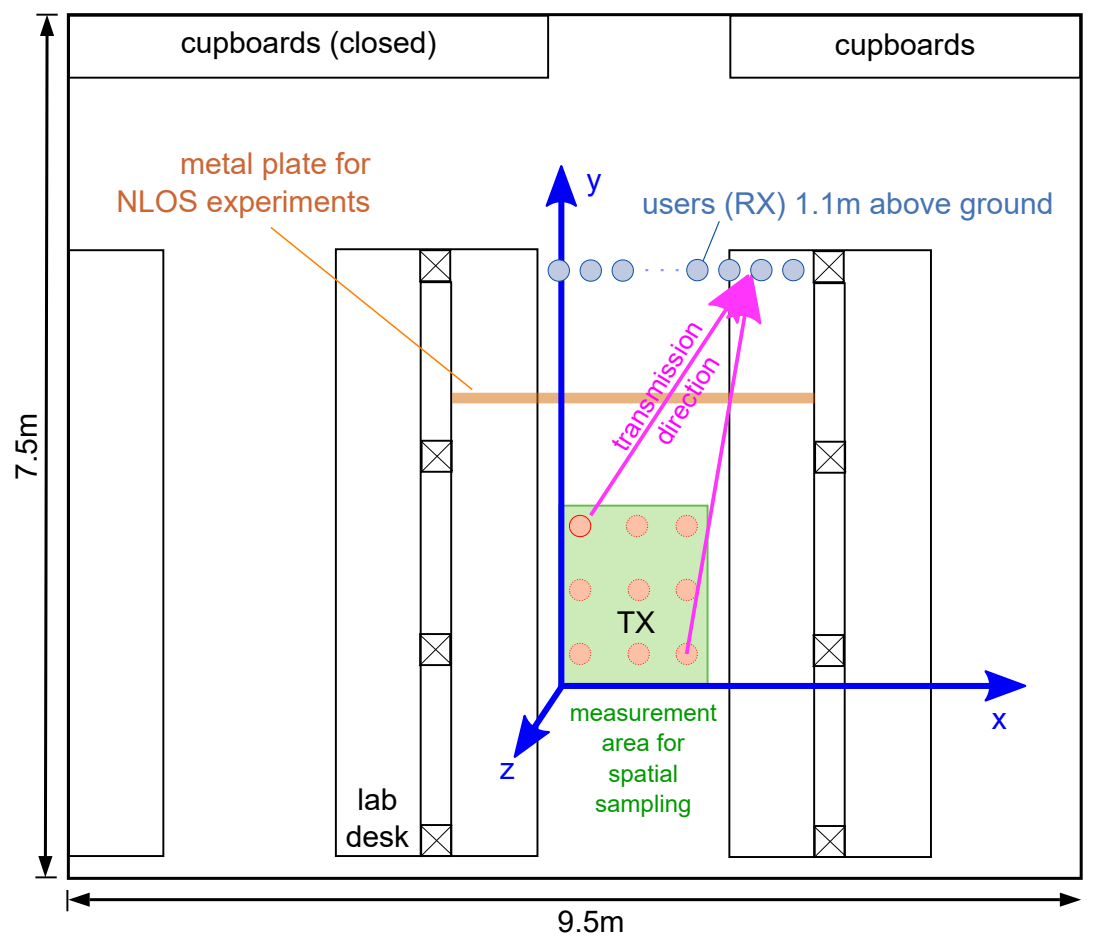

Fig. 9. Set-up for "inside out"-interpretation: Spatial samples of area measurement as a large antenna array to orthogonalize users outside.

Exploiting channel reciprocity, we can switch roles and interpret an arbitrary subset of spatial samples of the area measurement as basis for an antenna array of, e.g., an indoor access point; thus, the former "RX antennas" now form the access point antenna array, while the former "TX antennas" become up to 16 independent non-cooperating users (compare Fig. 9). By using a specific subset of the measured area points, we can set up different antenna array geometries, applied to orthogonalize the users outside.

For studying the achievable rates over the aperture size $A_{\text{eff}}$ a square antenna geometry of $3 \times 3$ antennas was cut out of the green area (Fig. 9), varying the distance $d$ between the antenna elements, with

$$d = \frac{\sqrt{A_{\text{eff}}}}{2}.$$

A precoding matrix $\mathbf{P}$ was used to precode to each of the $N = 16$ users. MR precoding was, again, chosen as a precoding scheme for this study. The used antenna geometry results in a channel matrix of dimensions $N \times M$, where $N$ is the number of virtual users and $M$ is the number of used area coordinate points (i.e., the antennas). With the MR precoding matrix $\mathbf{P}_{\text{MRC}}$ and using (2) the SIR for user $n$ can be calculated as

$$\text{SIR}_n = \frac{\left|\mathbf{h}_n \mathbf{h}_n^{\text{H}}\right|^2}{\sum\limits_{\substack{k=1 \\ k \neq n}}^{N} \left|\mathbf{h}_k \mathbf{h}_n^{\text{H}}\right|^2} \qquad (4)$$

where $\mathbf{h}_k$ is the $1 \times M$-channel vector with entries $h_{mk}$, i.e., from area points $m$ to user $k$. With (4) the sum rate for a specific scenario can be computed as

$$R_{sum} = \sum_{n=1}^{N} \log_2\left(1 + \text{SIR}_n\right) \qquad (5)$$

Note that (5) holds for interference-limited systems only, as is the case in our setup.

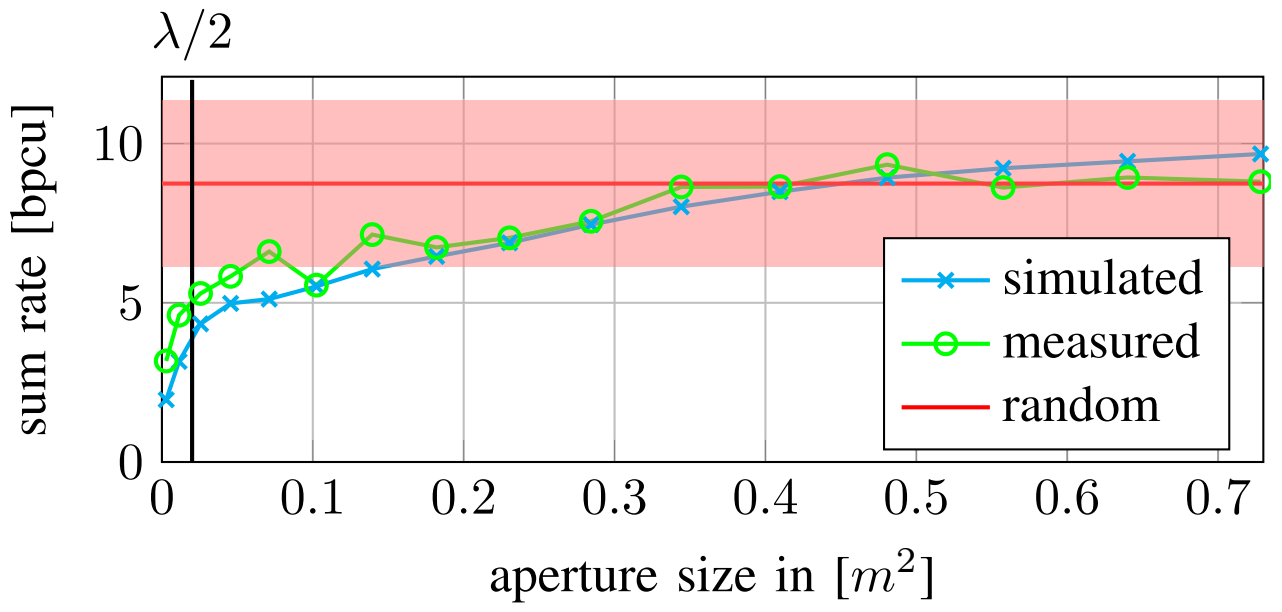

Fig. 10. Sum rate over aperture size in the LoS scenario for MR precoding and a square antenna geometry.

Fig. 10 shows the dependency between the aperture size and the sum rate for a fixed $3 \times 3$ square antenna geometry. The aperture size $A_{\text{eff}}$ at which the elements are at least $\lambda/2$ apart from each other is given as

$$A_{\text{eff}} = \left(2 \cdot \frac{\lambda}{2}\right)^2 \approx 0.02\text{m}^2.$$

Obviously, from Fig. 10, the biggest improvement in rate is obtained while the antenna aperture size is increased up to $A_{\text{eff}} \approx 0.02\text{m}^2$. Beyond that, only spatial diversity increases the rate. For reference, the red area indicates *that* range within which 90% of the rates using a *random* antenna geometry are located (with constraint that the 9 elements are at least $\lambda/2$ apart from each other). This verifies that, indeed, the aperture size is crucial for high achievable rates.

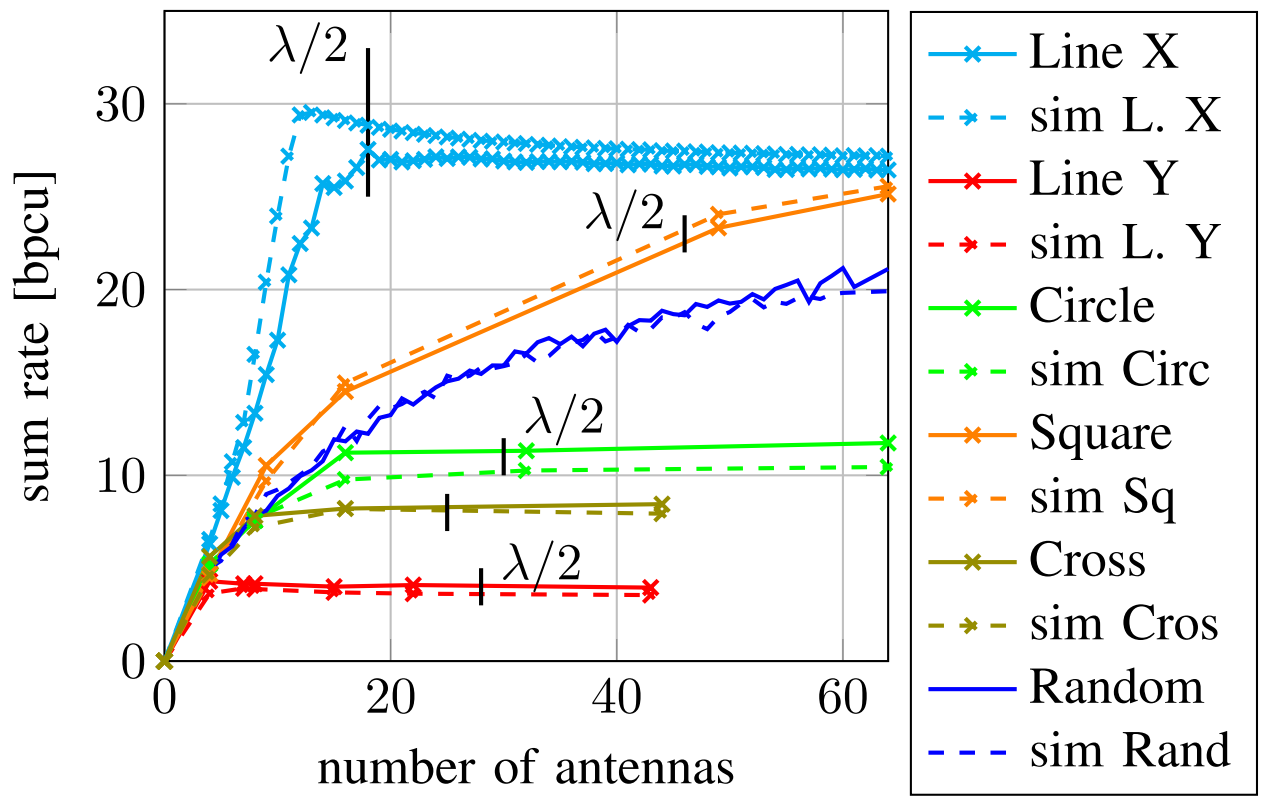

Fig. 11. Achievable rate vs number of used antennas in LoS scenario for MR precoding (simulation and measurement), different antenna geometries.

To find the best antenna array structure (independent of the aperture size) the measured area is used to create different array geometries. In Fig. 11 a line in $x$-direction, a line in $y$, a square, a cross and a circle array geometry are studied. The maximum aperture size within the measurement area was used for all antenna geometries. While keeping the aperture size fixed, the line in $y$ and the cross geometry could be evaluated up to 42 antennas, limited by the meander structure of the path the spider antenna was tracing (spacing of $4\,\text{cm}$ in $y$-direction). For reference, the dotted curves are based on simulation, as is possible for the LoS case. Note that all plotted curves are in good agreement with their corresponding simulated counterparts.

Furthermore, Fig. 11 shows that the array geometry of a line in $x$ has the highest achievable rate for all number of antennas, and the line in $y$ direction exhibits the lowest. Thus, only adding users in $x$-direction increases the overall rate, which is plausible since all users are arranged in parallel to the line in $x$-direction. For reference, the “random”-curve corresponds to a random antenna geometry with the constraint that the spacing between the antennas needs to be at least $\lambda/2$. Its rate falls right in between the $x$ and the $y$-line geometries. The square antenna geometry is better than the circle geometry as the users are, in this specific arrangement, on a line in front of the array, and not distributed on both sides of the array and therefore the created sidelobes are less important.

To further compare different precoding schemes, the NLoS scenario, as created by the tall metal plate inserted between TX and RX, is further investigated. Fig. 12 depicts sum rate results for different antenna geometries vs the number of antennas used, while applying MR precoding. The one-dimensional (i.e., linear) antenna array geometries “Line X, Y” capture the least amount of diversity, while the square array geometry benefits most from the available diversity: Its antenna elements are farthest apart and are distributed uniformly in $x$ and $y$- direction, respectively. Note that an array based on randomly distributed antenna elements performs almost as good as the square geometry.

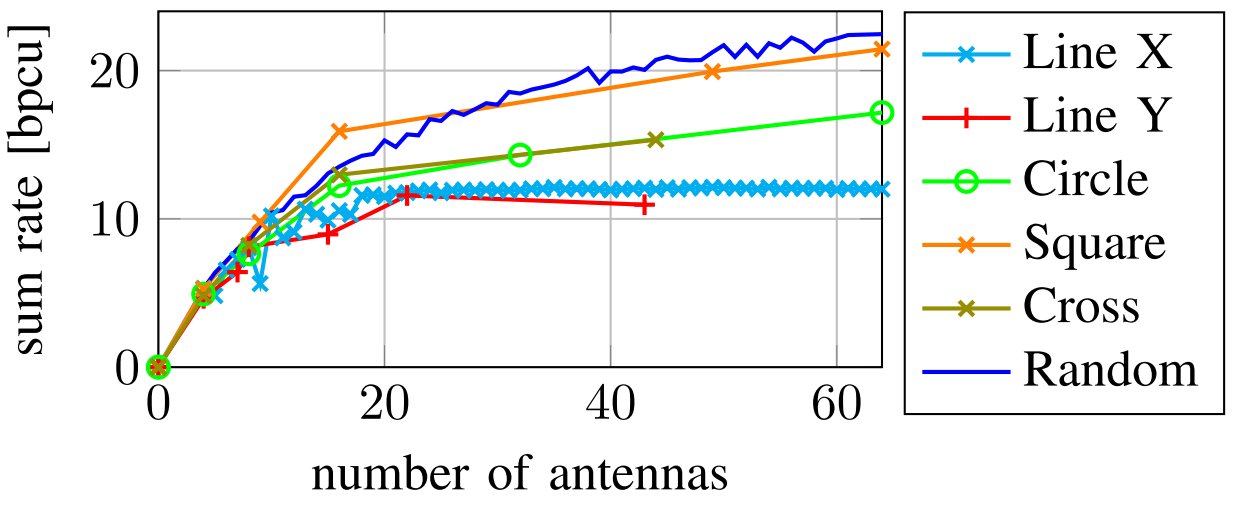

Fig. 12. Achievable rate vs number of used antennas in measured NLoS scenario for MR precoding, different antenna geometries.

To study different precoding schemes in the NLoS scenario, we now focus on the square antenna array geometry. For Phase Only (PO) precoding, the precoding matrix is given as

$$\mathbf{P}_{\text{PO}} = e^{j\cdot\arg(\mathbf{P}_{\text{MRC}})}$$

where the argument is taken element-by-element. The precoding matrix for Zero-Forcing (ZF) [12]

$$\mathbf{P}_{ZF} = \begin{cases} \sqrt{M-N}\cdot\left(\mathbf{H}^H\mathbf{H}\right)^{-1}\mathbf{H}^H & M > N \\ \sqrt{N-M}\cdot\mathbf{H}^H\left(\mathbf{H}\mathbf{H}^H\right)^{-1} & M \leq N \end{cases}$$

is calculated over the pseudo-inverse of the channel, forcing interference at the respective other 15 user positions to zero. Thus, the SIR for the precoding schemes can be calculated with (4) and (5), respectively.

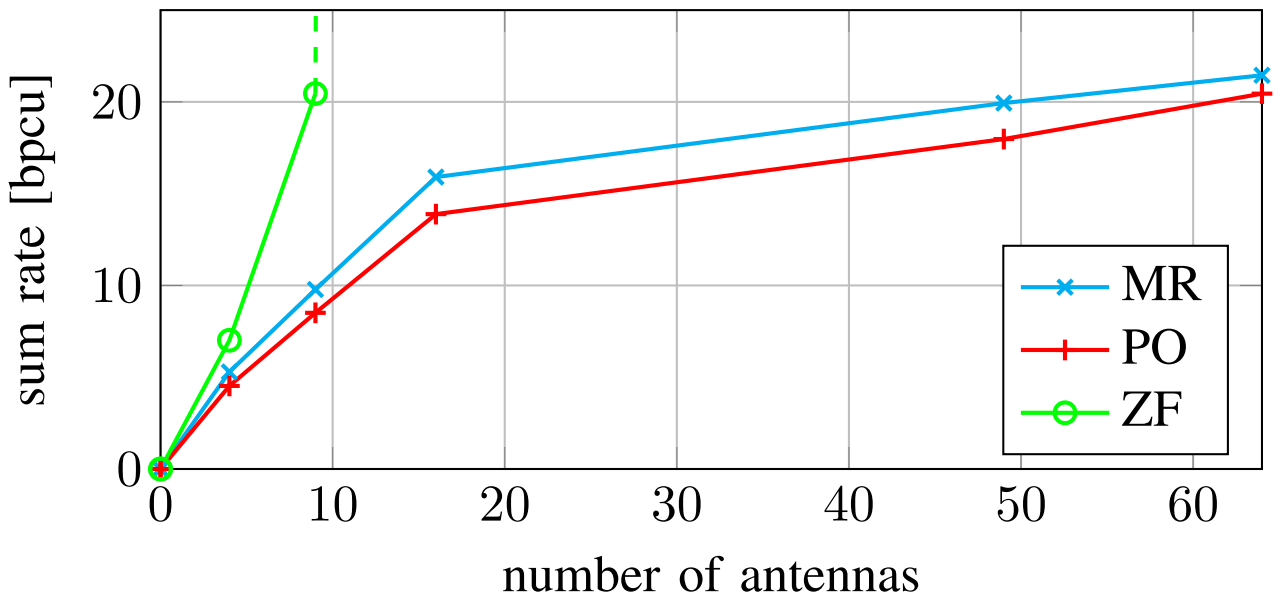

Fig. 13. Sum rate of MR, PO vs ZF-precoding in the measured NLoS scenario for a square antenna array geometry.

The comparison of the three different precoding schemes typically considered for large MIMO systems is shown in Fig. 13. As anticipated, ZF has the best performance; observe that, when increasing the number of antennas beyond 16, the linear equation system becomes overdetermined and enables to force the interference seen by the other users to zero. This results in very high rates (but not infinite rates since noise and bandwidth are now becoming the limiting factors). The MR precoding scheme performs better than the PO as it can also control the amplitudes of the TX antennas. Yet, this advantage is not very pronounced and only leads to small rate improvements. Thus, for its simplicity, PO precoding may become the preferred choice in practical systems.

## V. Conclusions

A spider antenna set-up for area/volume channel measurements was presented and verified by comparing against theoretical results in the LoS case for single and multi-antenna scenarios. Some possible applications were introduced, ranging from studying antenna geometries to user orthogonality for cluster modeling. A novel spatial SIR measure was used to better understand the transition from LoS to NLoS channel scenarios, enabling refined parameter definitions in future massive MIMO channel models. Moreover, the performance of different precoding schemes was compared for actually measured LoS and NLoS scenarios, showing that PO precoding does not lose much over MR precoding, and thus may be a pragmatic choice for future practical systems.